\documentclass[a4paper]{jpconf}
\usepackage{graphicx}
\begin{document}
\title{Schwinger pair creation of particles and strings}

\author{Christian Schubert}

\address{Instituto de F\'{\i}sica y Matem\'aticas,
Universidad Michoacana de San Nicol\'as de Hidalgo,
Edificio C-3, Apdo. Postal 2-82,
C.P. 58040, Morelia, Michoac\'an, M\'exico.}

\ead{schubert@ifm.umich.mx}

\begin{abstract}
I shortly review the worldline instanton method for calculating Schwinger
pair production rates in (i) one-loop QED (ii) multiloop QED and (iii) one-loop open string theory.
\end{abstract}

\newcommand{\bea}{\begin{eqnarray}}
\newcommand{\eea}{\end{eqnarray}}

\def\eqa{\!\!&=&\!\!}
\def\ccr{\nonumber\\}

\def\abs#1{\left| #1\right|}
\def\la{\langle}
\def\ra{\rangle}

\def\del{\Delta}
\def\ddel{{}^\bullet\! \Delta}
\def\deld{\Delta^{\hskip -.5mm \bullet}}
\def\ddeld{{}^{\bullet}\! \Delta^{\hskip -.5mm \bullet}}
\def\dddel{{}^{\bullet \bullet} \! \Delta}

\newcommand{\be}{\begin{equation}}
\newcommand{\ee}{\end{equation}}
\newcommand{\bear}{\begin{eqnarray}}
\newcommand{\ear}{\end{eqnarray}\noindent}
\newcommand{\no}{\noindent}
\newcommand{\veject}{\vfill\eject}
\newcommand{\ven}{\vfill\eject\noindent}
\date{}
\renewcommand{\theequation}{\arabic{section}.\arabic{equation}}
\renewcommand{\arraystretch}{2.5}
\newcommand{\GeV}{\mbox{GeV}}
\newcommand{\cL}{\cal L}
\newcommand{\D}{\cal D}
\newcommand{\Dhalf}{{D\over 2}}
\newcommand{\Det}{{\rm Det}}
\newcommand{\PP}{\cal P}
\newcommand{\G}{{\cal G}}
\def\R{1\!\!{\rm R}}
\def\Eins{\mathord{1\hskip -1.5pt
\vrule width .5pt height 7.75pt depth -.2pt \hskip -1.2pt
\vrule width 2.5pt height .3pt depth -.05pt \hskip 1.5pt}}
\newcommand{\symb}{\mbox{symb}}
\renewcommand{\arraystretch}{2.5}
\newcommand{\slD}{\raise.15ex\hbox{$/$}\kern-.57em\hbox{$D$}}
\newcommand{\slpartial}{\raise.15ex\hbox{$/$}\kern-.57em\hbox{$\partial$}}
\newcommand{\slG}{{{\dot G}\!\!\!\! \raise.15ex\hbox {/}}}
\newcommand{\Gd}{{\dot G}}
\newcommand{\Gund}{{\underline{\dot G}}}
\newcommand{\Gdd}{{\ddot G}}
\def\slash#1{#1\!\!\!\raise.15ex\hbox {/}}
\def\GBd12{{\dot G}_{B12}}
\def\mneg{\!\!\!\!\!\!\!\!\!\!}
\def\Mneg{\!\!\!\!\!\!\!\!\!\!\!\!\!\!\!\!\!\!\!\!}
\def\exmn{\Bigl(\mu \leftrightarrow \nu \Bigr)}
\def\non{\nonumber}
\def\beqn*{\begin{eqnarray*}}
\def\eqn*{\end{eqnarray*}}
\def\sy{\scriptscriptstyle}
\def\footstrut{\baselineskip 12pt}
\def\square{\kern1pt\vbox{\hrule height 1.2pt\hbox{\vrule width 1.2pt
   \hskip 3pt\vbox{\vskip 6pt}\hskip 3pt\vrule width 0.6pt}
   \hrule height 0.6pt}\kern1pt}
\def\slash#1{#1\!\!\!\raise.15ex\hbox {/}}
\def\dint#1{\int\!\!\!\!\!\int\limits_{\!\!#1}}
\def\bra#1{\langle #1 |}
\def\ket#1{| #1 \rangle}
\def\vev#1{\langle #1 \rangle}
\def\rightvac{\mid 0\rangle}
\def\leftvac{\langle 0\mid}
\def\dps{\displaystyle}
\def\sy{\scriptscriptstyle}
\def\half{{1\over 2}}
\def\third{{1\over3}}
\def\fourth{{1\over4}}
\def\fifth{{1\over5}}
\def\sixth{{1\over6}}
\def\seventh{{1\over7}}
\def\eigth{{1\over8}}
\def\ninth{{1\over9}}
\def\tenth{{1\over10}}
\def\pa{\partial}
\def\ddtau{{d\over d\tau}}
\def\ge{\hbox{\textfont1=\tame $\gamma_1$}}
\def\gz{\hbox{\textfont1=\tame $\gamma_2$}}
\def\gd{\hbox{\textfont1=\tame $\gamma_3$}}
\def\go{\hbox{\textfont1=\tamt $\gamma_1$}}
\def\gt{\hbox{\textfont1=\tamt $\gamma_2$}}
\def\gth{\hbox{\textfont1=\tamt $\gamma_3$}} 
\def\gf{\hbox{$\gamma_5\;$}}
\def\ie{\hbox{$\textstyle{\int_1}$}}
\def\iz{\hbox{$\textstyle{\int_2}$}}
\def\id{\hbox{$\textstyle{\int_3}$}}
\def\ldop{\hbox{$\lbrace\mskip -4.5mu\mid$}}
\def\rdop{\hbox{$\mid\mskip -4.3mu\rbrace$}}
\def\eps{\epsilon}
\def\epshalf{{\epsilon\over 2}}
\def\e{\mbox{e}}
\def\g{\mbox{g}}
\def\kinb{{1\over 4}\dot x^2}
\def\kinf{{1\over 2}\psi\dot\psi}
\def\expk{{\rm exp}\biggl[\,\sum_{i<j=1}^4 G_{Bij}k_i\cdot k_j\biggr]}
\def\expp{{\rm exp}\biggl[\,\sum_{i<j=1}^4 G_{Bij}p_i\cdot p_j\biggr]}
\def\expshort{{\e}^{\half G_{Bij}k_i\cdot k_j}}
\def\expabb{{\e}^{(\cdot )}}
\def\epseps#1#2{\varepsilon_{#1}\cdot \varepsilon_{#2}}
\def\epsk#1#2{\varepsilon_{#1}\cdot k_{#2}}
\def\kk#1#2{k_{#1}\cdot k_{#2}}
\def\G#1#2{G_{B#1#2}}
\def\Gp#1#2{{\dot G_{B#1#2}}}
\def\GF#1#2{G_{F#1#2}}
\def\Dab{{(x_a-x_b)}}
\def\Dsq{{({(x_a-x_b)}^2)}}
\def\lag{( -\partial^2 + V)}
\def\PITD{{(4\pi T)}^{-{D\over 2}}}
\def\4piTD{{(4\pi T)}^{-{D\over 2}}}
\def\4piT4{{(4\pi T)}^{-2}}
\def\TintmD{{\dps\int_{0}^{\infty}}{dT\over T}\,e^{-m^2T}
    {(4\pi T)}^{-{D\over 2}}}
\def\go{\hbox{\textfont1=\tamt $\gamma_1$}}
\def\gt{\hbox{\textfont1=\tamt $\gamma_2$}}
\def\gth{\hbox{\textfont1=\tamt $\gamma_3$}} 
\def\gf{\hbox{$\gamma_5\;$}}
\def\ie{\hbox{$\textstyle{\int_1}$}}
\def\iz{\hbox{$\textstyle{\int_2}$}}
\def\id{\hbox{$\textstyle{\int_3}$}}
\def\ldop{\hbox{$\lbrace\mskip -4.5mu\mid$}}
\def\rdop{\hbox{$\mid\mskip -4.3mu\rbrace$}}
\def\eps{\epsilon}
\def\epshalf{{\epsilon\over 2}}
\def\e{\mbox{e}}
\def\g{\mbox{g}}
\def\kinb{{1\over 4}\dot x^2}
\def\kinf{{1\over 2}\psi\dot\psi}
\def\expk{{\rm exp}\biggl[\,\sum_{i<j=1}^4 G_{Bij}k_i\cdot k_j\biggr]}
\def\expp{{\rm exp}\biggl[\,\sum_{i<j=1}^4 G_{Bij}p_i\cdot p_j\biggr]}
\def\expshort{{\e}^{\half G_{Bij}k_i\cdot k_j}}
\def\expabb{{\e}^{(\cdot )}}
\def\epseps#1#2{\varepsilon_{#1}\cdot \varepsilon_{#2}}
\def\epsk#1#2{\varepsilon_{#1}\cdot k_{#2}}
\def\kk#1#2{k_{#1}\cdot k_{#2}}
\def\G#1#2{G_{B#1#2}}
\def\Gp#1#2{{\dot G_{B#1#2}}}
\def\GF#1#2{G_{F#1#2}}
\def\Dab{{(x_a-x_b)}}
\def\Dsq{{({(x_a-x_b)}^2)}}
\def\lag{( -\partial^2 + V)}
\def\PITD{{(4\pi T)}^{-{D\over 2}}}
\def\4piTD{{(4\pi T)}^{-{D\over 2}}}
\def\4piT4{{(4\pi T)}^{-2}}
\def\TintmD{{\dps\int_{0}^{\infty}}{dT\over T}\,e^{-m^2T}
    {(4\pi T)}^{-{D\over 2}}}
\def\Tintm4{{\dps\int_{0}^{\infty}}{dT\over T}\,e^{-m^2T}
    {(4\pi T)}^{-2}}
\def\Tintm{{\dps\int_{0}^{\infty}}{dT\over T}\,e^{-m^2T}}
\def\Tint{{\dps\int_{0}^{\infty}}{dT\over T}}
\def\pint{{\dps\int}{dp_i\over {(2\pi)}^d}}
\def\Dx{\dps\int{\cal D}x}
\def\Dy{\dps\int{\cal D}y}
\def\Dpsi{\dps\int{\cal D}\psi}
\def\Tr{{\rm Tr}\,}
\def\tr{{\rm tr}\,}
\def\sumij{\sum_{i<j}}
\def\freeexp{{\rm e}^{-\int_0^Td\tau {1\over 4}\dot x^2}}
\def\arraystretch{2.5}
\def\Ge{\mbox{GeV}}
\def\dA{\partial^2}
\def\DA{\sqsubset\!\!\!\!\sqsupset}
\def\FFdual{F\cdot\tilde F}
\def\mn{\mu\nu} 

\pagestyle{empty}

\renewcommand{\thefootnote}{\fnsymbol{footnote}}


\centerline{Talk given at {\it  XIV Mexican School of Particles and Fields}, November 8 - 12, 2010, Morelia, Mexico}

\section{History}

\renewcommand{\theequation}{1.\arabic{equation}}
\setcounter{equation}{0}

\begin{figure}[h]
\hspace{100pt}{\centering
\includegraphics{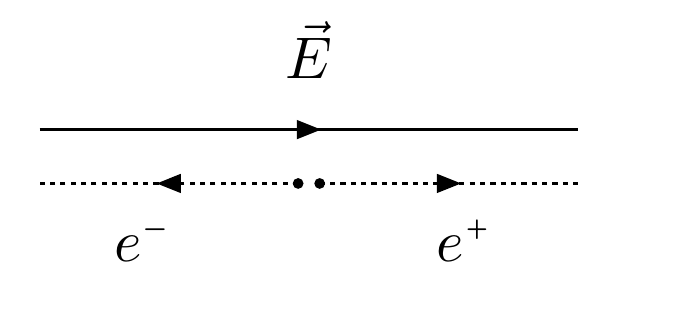}
}
\caption{Pair creation by an external field.}
\label{fig1}
\end{figure}

Already in 1931 F. Sauter \cite{sauter} realized that Dirac's theory of the electron predicts that an electric field
of sufficient strength and extent can induce spontaneous creation of electron -- positron pairs from the vacuum. 
Naively, a virtual pair turns real by separating out
along the field and drawing a sufficient amount of energy from it to make up for both rest mass energies (fig. \ref{fig1}).

\bigskip

Twenty years later, J. Schwinger \cite{schwinger51} used the more advanced methods
then available to obtain his famous formulas for the
imaginary parts of the one-loop effective actions in a constant homogeneous
field, for both spinor and scalar QED:

\bear
{\rm Im} {\cal L}^{(1)}_{\rm spin}(E) &=&  \frac{m^4}{8\pi^3}
\beta^2\, \sum_{n=1}^\infty \frac{1}{n^2}
\,\exp\left[-\frac{\pi n}{\beta}\right]
\label{schwingerspin}\\
{\rm Im} {\cal L}_{\rm scal}^{(1)}(E) &=& - \frac{m^4}{16\pi^3}
\beta^2\, \sum_{n=1}^\infty \frac{(-1)^n}{n^2}
\,\exp\left[-\frac{\pi n}{\beta}\right]
\label{schwingerscal}
\ear
($\beta = {eE\over m^2}$; the upper index on $\cal L$ refers to the loop order).
For low pair creation rates, these imaginary parts relate directly to the pair creation rate per volume $w$,
namely one has $w \approx 2 {\rm Im}{\cal L}$. 
Schwinger's formulas (\ref{schwingerspin}),(\ref{schwingerscal}) are written in terms of an infinite sum of ``Schwinger exponentials'', where
the $n$th term relates to the coherent production of $n$ pairs by the field. 
Their dependence on the field strength is nonperturbative, which confirms Sauter's picture
of the pair production as a tunneling process. 
Unfortunately, this also implies that the pair creation rate is 
exponentially small for $E  \ll E_{\rm cr} := 10^{16} V/{\rm cm}$.
Until recently, macroscopic fields of this strength were considered unattainable.
However, this has changed due to the rapid evolution of laser technology. 
Presently there are various lasers in planning or operation, e.g. POLARIS \cite{polaris},
ELI \cite{eli} and XFEL \cite{xfel}, whose maximum 
field strength falls only a few orders of magnitude short of $E_{\rm cr}$ 
(only about one order of magnitude in the case of XFEL).
This has led to a number of attempts to bring down the pair creation
threshold by a superposition of various optical or X-ray beams; to mention
just two recent proposals, counterpropagating linearly polarized lasers
were considered in \cite{rufetal} and the superposition of a plane-wave
X-ray beam with a strongly focused optical laser pulse in \cite{dugisc}.
Obviously, such configurations have field strengths very far from the
constant homogeneous case considered by Schwinger, and there is
little hope for an exact calculation of the associated pair creation rates;
approximation methods will have to be used. 
The tunnel effect picture suggest the use of WKB, and this is indeed
the method which in the past has been almost universally used
for the calculation of pair creation rates in nonconstant fields;
see, e.g., \cite{keldysh65,breitz70,narnik70,popov72,popmar72}.

The subject of this talk, the {\sl worldline instanton method}, is 
related to the WKB approximation, but closer in spirit to modern relativistic quantum
field theory, which leads to certain advantages. This approach was invented
in 1982 by I. K. Affleck, O. Alvarez and N. S. Manton \cite{afalma} (``AAM'' in the following)
for the case of a constant field in scalar QED, using Feynman's worldline representation of the scalar QED effective action
\cite{feynman50}. In the quenched (single scalar loop) approximation, this representation is

\bear
\Gamma_{\rm scal}[A] &=&
\int d^4x\, {\cal L}(A) =
\int_0^{\infty}{dT\over T}\,{\rm e}^{-m^2T}
{\displaystyle \int_{x(T)=x(0)}}{\cal D}x(\tau)
\, e^{-S[x(\tau)]}
\label{Gammascal}
\ear
Here $m$ and $T $ are the mass and proper time of the loop scalar,
and the path integral $\int{\cal D}x(\tau)$
is over closed trajectories in (Euclidean) spacetime with a worldline action
$\ S[x(\tau)]$ that has three parts:

\bear
S&=& S_0 + S_{\rm ext} + S_{\rm int}
\label{decompS}
\ear
where

\bear
S_0 &=& \int_0^T d\tau {\dot x^2\over 4} \nonumber\\
S_{\rm ext} &=& ie\int_0^T \dot x^{\mu}A_{\mu}(x(\tau)) 
\nonumber\\
S_{\rm int} &=&
-{e^2\over 8\pi^2}\int_0^Td\tau_1\int_0^Td\tau_2 {\dot x(\tau_1)\cdot\dot x(\tau_2)\over
(x(\tau_1)-x(\tau_2))^2}
\label{Sparts}
\ear
$S_0$ describes the free propagation of the scalar particle,
$S_{\rm ext}$ its interaction with the background field,
and $S_{\rm int}$ internal photon exchanges in the loop.

The basic idea is to calculate the path integral by a stationary phase approximation.
We will carry it through first at the one-loop level.

\section{One-loop QED}

\renewcommand{\theequation}{2.\arabic{equation}}
\setcounter{equation}{0}

The one-loop effective action is obtained from (\ref{Gammascal}) by omitting $S_{\rm int}$,

\bear
\Gamma_{\rm scal}^{(1)} [A] &=&
\int_0^{\infty}{dT\over T}\, \e^{-{m^2}T}
\int {\cal D}x 
\, \e^{-\int_0^Td\tau 
\bigl({\dot x^2\over 4} +ieA\cdot \dot x \bigr)}
\label{Gammascal1loop}
\ear
Rescaling $\tau = Tu$, this becomes
\vspace{-5pt}
\bear
\Gamma_{\rm scal}^{(1)} [A] &=&
\int_0^{\infty}{dT\over T}\, \e^{-{m^2}T}
\int {\cal D}x 
\, \e^{-\Bigl({1\over T}\int_0^1du \,
\dot x^2 +ie\int_0^1du A\cdot \dot x 
\Bigr) }
\label{Gammascal1looprescaled}
\ear
This makes it apparent that the $T$ integral has a
stationary point at 
$T_c = \sqrt{\int_0^1 du\,\dot x^2}/m$.
If we are only interested in the imaginary part of the effective
action at large mass, 
we can use this stationary point to eliminate the $T$ integral, yielding

\bear
{\rm Im} \Gamma^{(1)}_{\rm scal}[A] &=&
{1\over m}\sqrt{2\pi\over T_c}
\,{\rm Im} \int {\cal D}x \, 
\e^{-\Bigl(m\sqrt{\int du \,\dot x^2} 
+ie\int_0^1 du A\cdot \dot x
\Bigr)}
\label{Gammascal1loopIm}
\ear
The new worldline action,

\vspace{-15pt}

\bear
S &=& m\sqrt{\int_0^1 du\, \dot x^2} + ie \int_0^1duA\cdot \dot x
\label{Snew}
\ear
is stationary if

\vspace{-18pt}

\bear
m{\ddot x_{\mu}\over \sqrt{\int_0^1 du\,\dot x^2}} &=& ie F_{\mn}\dot x_{\nu}
\label{eom}
\ear
Contracting this equation with $\dot x^{\mu}$ shows that
$\dot{x}^2={\rm const.}\equiv a^2$, so that
$m\ddot x_{\mu} = iea F_{\mn}\dot x_{\nu}$.
Thus the extremal action trajectory $\ x^{\rm cl}(u)$, to be called {\sl worldline instanton},
is simply a periodic solution of the Lorentz  force equation.
The semiclassical (large mass) approximation becomes 

\vspace{-15pt}

\bear
{\rm Im}{\cal L}(E)  
\,\,
\sim
\,\, {\rm e}^{-S[x^{\rm cl}]}
\label{approx}
\ear
In the case of a constant field ${\bf E} = (0,0,E)$ considered by AAM, the worldline instanton 
turns out to be a circle in the $x_3-x_4$ plane, of radius $m/eE$ and winding number $n$:

\vspace{-6pt}
\bear
x^{\rm cl}(u) &=& {m\over eE}\,\Bigl(x_1,x_2,{\rm cos}(2n\pi u),{\rm sin}(2n\pi u)\Bigr)\label{instconst}\\
S[x^{\rm cl}] &=& n\pi\, {m^2\over eE}\label{Sconst}
\ear
Thus the instanton action for winding number $n$ reproduces the $n$th exponent in Schwinger's
representation (\ref{schwingerscal}) of ${\rm Im}{\cal L}^{(1)}_{\rm scal}(E)$.

Worldline instantons for more general electric fields where obtained in \cite{wlinst1}. 
As an example, let us show the case of the ``single-bump'' field
$E(x_3)=E\, {\rm sech}^2(kx_3)$. Here the $n$th instanton solution is

\vspace{-15pt}

\bear
x_3(u)&=&
{m\over eE}{1\over\tilde\gamma}\,{\rm arcsinh}\biggl({\tilde\gamma\over
\sqrt{1-\tilde\gamma^2}}\,\sin(2n\pi u)\biggr)\non\\
x_4(u)&=&
{m\over eE}{1\over\tilde\gamma \sqrt{1-\tilde\gamma^2}}\,{\rm arcsin}
\bigl(\tilde\gamma \cos(2n\pi u)\bigr)
\label{instbump}
\ear
where the parameter $\tilde \gamma\equiv \frac{m k}{eE}$ is a measure for the inhomogeneity of the field.
The instanton action is

\begin{figure}[h]
\centerline{\includegraphics[scale=.8]{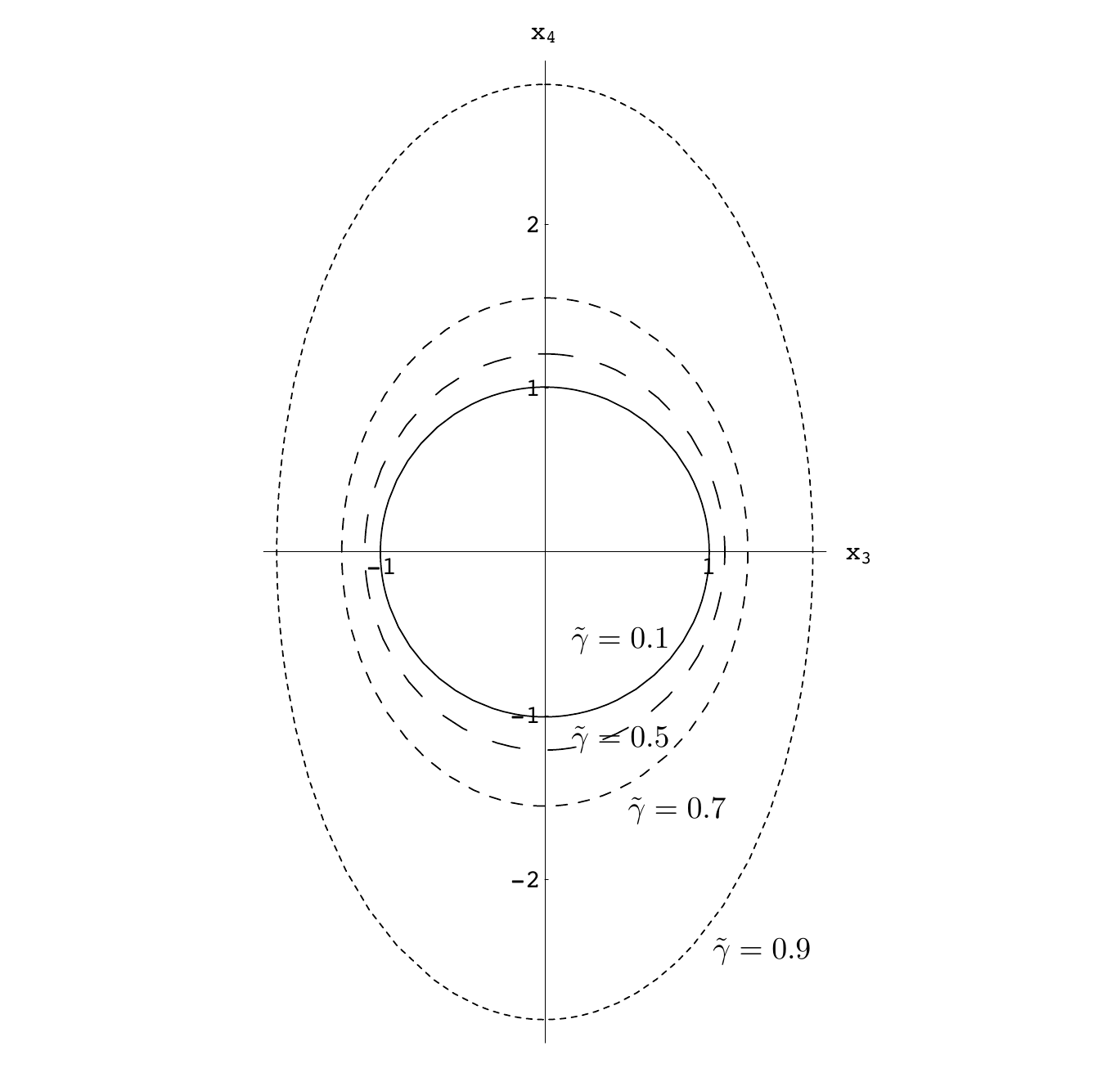}}
\caption{Plot of the instanton paths 
for the case $E(x)=E{\rm sech}^2(kx)$.}
\label{singlebump}
\end{figure}

\bear
S[x^{\rm cl}] &=&n\, \frac{m^2 \pi}{e E}\left(\frac{2}{1+\sqrt{1-\tilde\gamma^2}}\right)\label{Sbump}
\ear
For $\tilde\gamma \to 0$ one recovers the constant field instanton. With increasing 
$\tilde\gamma$ the instanton action increases, which means that the associated pair creation rate 
goes down. For $\tilde\gamma > 1$ the instanton ceases to exist, in accordance with the fact that the maximal energy which can
be extracted from the field by a virtual pair drops below $2m$. In fig. \ref{singlebump} the instanton trajectories are plotted for
various values of $\tilde\gamma$.

$\phantom{r}$

In \cite{wlinst2} a general method was developed to also
calculate the prefactor of the instanton-induced Schwinger exponential $\e^{-S[x^{\rm cl}]}$, that is,
the fluctuation determinant of the expansion of the worldline path integral around the stationary trajectory.
Including this prefactor for the single-bump case one finds excellent numerical agreement \cite{wlinst2} 
with an exact result by A.I. Nikishov \cite{nikishov} and as well with a direct Monte Carlo integration 
of the worldline path integral \cite{giekli}.

See \cite{wlinst2} for some other cases where the instantons can be found in closed form,  {\cite{dunwan} for  
more general classes of fields, and \cite{wlinst1} for the extension to spinor QED. 

\section{Multiloop QED}

\renewcommand{\theequation}{3.\arabic{equation}}
\setcounter{equation}{0}

As argued already by \cite{afalma}, the constant field
worldline instantons (\ref{instconst}) remain stationary trajectories even in the presence of photon
insertions. Evaluating the photon insertion term $S_{int}$ on the principal ($n=1$)  trajectory gives
the following all-loop formula for the imaginary part of the effective Lagrangian in the
large mass (= weak field) approximation:

\bear
{\rm Im}{\cal L}_{\rm scal}(E) = \sum_{l=1}^{\infty}{\rm Im}{\cal L}^{(l)}_{\rm scal}(E)
&{\stackrel{\beta\to 0}{\sim}}&
 \frac{m^4\beta^2}{8\pi^3}
\,{\rm exp}\Bigl[ -{\pi\over\beta}+\alpha\pi \Bigr]
=
{\cal L}_{\rm scal}^{(1)}\,\e^{\alpha\pi}
\label{AAM}
\ear
This formula is rather remarkable, since in terms of standard field theory it corresponds to a summation
of the infinite set of Feynman diagrams shown in fig. \ref{AAMfeyn}:

\vspace{20pt}

\begin{figure}[h]
{\centering
\includegraphics{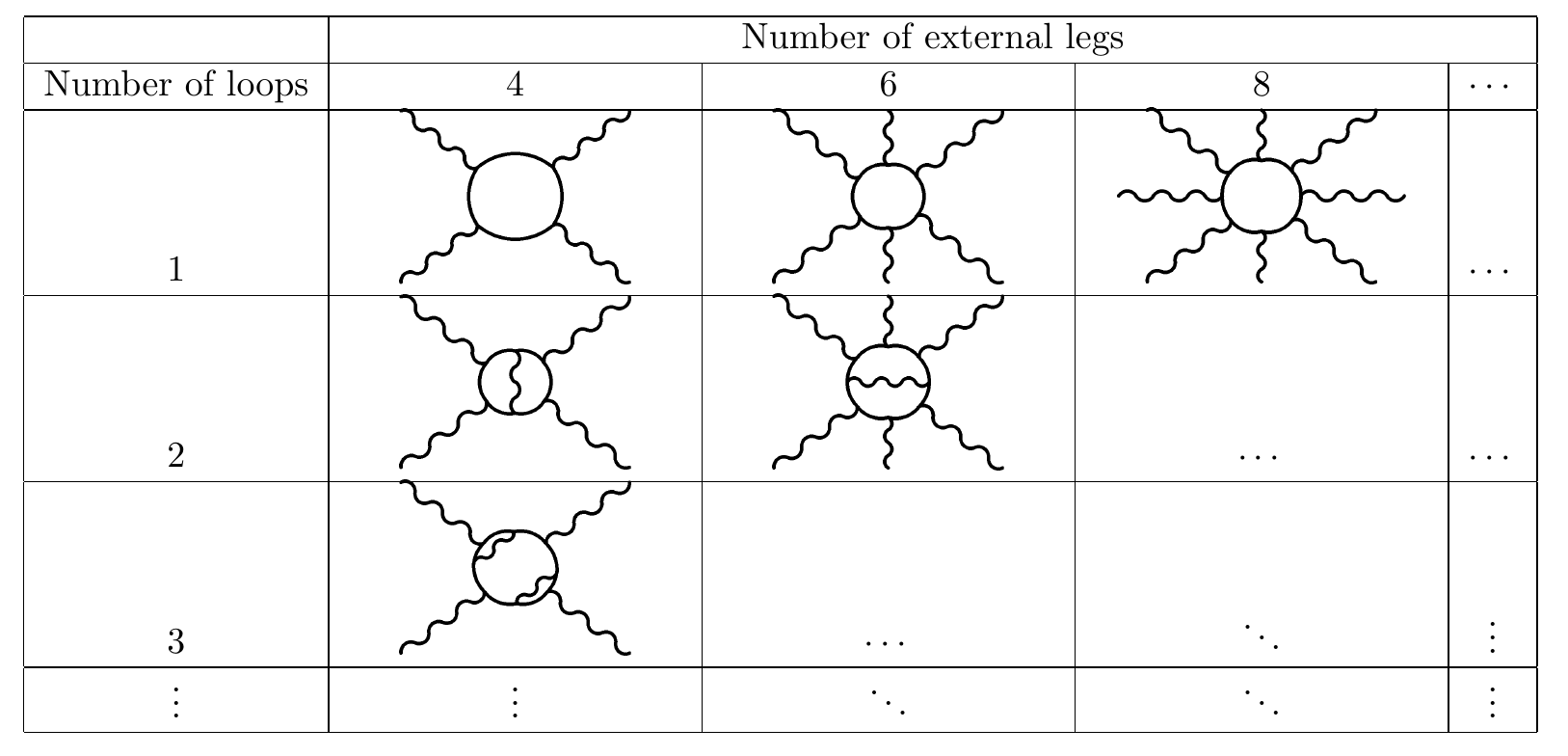}
}
\caption{Feynman diagrams contributing to the AAM formula.}
\label{AAMfeyn}
\end{figure}

\no
Moreover, the mass in (\ref{AAM}) is the physical mass \cite{afalma}, so that
fig. \ref{AAMfeyn} effectively includes also an infinite set of
mass renormalization counterdiagrams.

In 1984, S. L. Lebedev and V. I. Ritus \cite{lebrit,ginzburg} 
obtained the same exponentiation by an analysis 
of the corrections due to photon exchanges in Sauter's tunnelling picture,
for both scalar and spinor QED. They also confirmed
this $\e^{\alpha\pi}$ factor to linear order in $\alpha$ by
a direct calculation of ${\cal L}_{\rm spin}^{(2)}(E)$. 

Apart from the Schwinger pair creation effect,
the AAM formula (\ref{AAM}) and its spinor QED equivalent hold also
interesting information on the multiloop QED photon amplitudes, in the
limit of low photon energy and large photon number. This information can be extracted using
Borel dispersion relations \cite{dunsch1,dunschSD2}, which allows one to show, for example, that the
large $N$ behaviour of these amplitudes for $l\geq 2$ is
dominated by the one-fermion loop contributions, and
qualitatively different for physical and generic mass renormalization.
More interestingly, it suggests that the QED
perturbation series may converge for the one-fermion loop contributions
to the $N$ -- photon amplitudes  \cite{dunschSD2,colima,mascvi,humcsc}.
If correct, this would extend an old conjecture by P. Cvitanovic \cite{cvitanovic} for the
$g-2$ factor.
 
\section{One-loop open string theory}

\renewcommand{\theequation}{4.\arabic{equation}}
\setcounter{equation}{0}

At the one-loop level, pair creation by a constant homogeneous electric field has also been studied for string theory,
both for the open \cite{burgess,bacpor} and the heterotic string \cite{bacpor}.
For the open string case, C. Bachas and M. Porrati \cite{bacpor} obtained the following
formula for the imaginary part of the effective action:

\bear
&&{\rm Im}{\cal L}_{\rm string}^{(1)}(E) = \frac{1}{4(2\pi)^{D-1}}\sum_{\rm states \,
S}\frac{\beta_1 + \beta_2}{\pi \epsilon}
\sum_{n=1}^{\infty}(-)^{n+1}
\Bigl(\frac{\abs{\epsilon}}{n}\Bigr)^{D/2}
\,{\rm exp}\Bigl(-\frac{\pi
n}{\abs{\epsilon}}(M_S^2+\epsilon^2)\Bigr)
\nonumber\\
\label{bachasporrati}
\ear
Here the first sum is over the physical states of the bosonic string, with
$M_S$ the mass
of the state. The second sum is a Schwinger-type sum. $D=26$ is the critical
dimension for the open string. We have defined

\bear
\beta_{1,2} &=& \pi q_{1,2} E
\label{defbeta}
\ear
where $q_{1,2}$ are the $U(1)$ charges at the string endpoints, and

\bear
\epsilon &=& \frac{1}{\pi} \Bigl({\rm arctanh}\beta_1+{\rm
arctanh}\beta_2\Bigr)
\label{eps}
\ear
In the weak field limit, (\ref{bachasporrati})  reproduces the scalar Schwinger
formula (\ref{schwingerscal}), as well as its generalization to arbitrary integer spin $J$.
For stronger fields the right hand grows much faster than for the particle case, and even 
diverges at a critical field strength 

\bear
E_{\rm cr} &=& \frac{1}{\pi \abs{{\rm max} \,q_i}}
\label{Ecr}
\ear\
Presumably this means that a field of this strength would break the string apart.

We will now show how to reproduce, in the large mass limit, the Schwinger exponents in the open-string result of (\ref{bachasporrati})  
by a worldsheet instanton which is a natural generalization of the AAM worldline instanton (\ref{instconst})
\cite{schtor}.

To write down a path integral for  
the open string at one-loop in an external Maxwell field, one has to replace the particle loop by a string annulus
and insert the interaction term $S_{\rm ext}$ of (\ref{Sparts}) on both boundaries of the loop, with the charge $e$ replaced by 
$q_{1,2}$ (we assume $q_1\ne q_2$ to exclude the M\"obius strip contribution to the amplitude).
The effective action becomes, in conformal gauge,

\bear
\Gamma_{\rm string}^{(1)}[F] &=& \half
\int_0^{\infty}{dT\over T}
(4\pi^2 T)^{-{D\over 2}}
Z(T)
\int {\cal D}x \,\e^{-S_E[x,F]}
\label{Gammastring}
\ear
\no
where the path integral is over the embeddings of the annulus at fixed $T$ into
$D$ - dimensional flat space. 
The worldsheet action is

\bear
S_E  &=& 
\frac{1}{4\pi\alpha'}\int d\sigma d\tau \,\partial_a x^{\mu} \partial^a x_{\mu}
-i {q_1\over 2} \int d\tau\,
x^{\mu}\partial_{\tau}x^{\nu}F_{\mu\nu}\Bigr\vert_{
\sigma =0}
-i {q_2\over 2} \int d\tau\,
x^{\mu}\partial_{\tau}x^{\nu}F_{\mu\nu}\Bigr\vert_{\sigma = \half}
\nonumber\\
\label{SE}
\ear
Here $\alpha'$ is the Regge slope, which will be set equal to
$\half$ in the following. The worldsheet is parametrized as a rectangle
$\sigma \in \lbrack 0, \half\rbrack$ and $\tau \in \lbrack 0 , T\rbrack$ where
$\tau = T$ is identified with $\tau = 0$. 
We use euclidean conventions
where $\sigma^0 = -i\sigma^2 = -i\tau $, $x^0 = -ix^D$, and $A_D = - iA_0$.
$Z(T)$ is the partition function, 

\bear
Z(T) &=& \sum_{\rm oriented\,\, states} \e^{-\pi T M_S^2}
\label{Z}
\ear
Let us now consider the constant electric field case,
$F_{D,D-1}=-F_{D-1,D} = iE$.
We rescale $\tau = Tu$ and do the $T$ - integral by the method of steepest
descent. The stationary point is

\bear
T_0 &=& \sqrt{\frac{I_u}{I_{\sigma} + 2\pi^2M_S^2}}
 \label{T0}
\ear
where

\bear
I_{\sigma} &:=& \int_0^1du\int_0^{\half}d\sigma \,\partial_{\sigma} x^{\mu}
\partial_{\sigma} x_{\mu} \nonumber\\
I_{u} &:=& \int_0^1du\int_0^{\half}d\sigma \,\partial_u x^{\mu} \partial_u
x_{\mu} 
\label{Isigma}
\ear
The new worldsheet action is

\bear
S_{\rm eff} &=& \frac{1}{\pi} \sqrt{I_u}\sqrt{I_{\sigma}+2\pi^2M_S^2}
-i {q_1\over 2} \int d\tau\,
x^{\mu}\partial_{\tau}x^{\nu}F_{\mu\nu}\Bigr\vert_{
\sigma =0}
 -i {q_2\over 2} \int d\tau\,
x^{\mu}\partial_{\tau}x^{\nu}F_{\mu\nu}\Bigr\vert_{\sigma = \half}
\nonumber\\
\label{Seff}
\ear
It leads to the equations of motion

\bear
\Bigl\lbrack I_u\partial_{\sigma}^2 + (I_{\sigma}+2\pi^2M_S^2)
\partial_{u}^2\Bigr\rbrack\, x^{\mu} &=& 0 \nonumber\\
T_0\partial_{\sigma}x^{\mu} &=& i\pi q_2 F_{\mn}\partial_{u}x^{\nu}\,\,
\qquad (\sigma = \half) \nonumber\\
T_0\partial_{\sigma}x^{\mu} &=& - i\pi q_1 F_{\mn}\partial_{u}x^{\nu}
\qquad (\sigma = 0) \nonumber\\
\label{eomstring}
\ear
The $n$th worldsheet instanton solving these equations is 
found by the ansatz

\bear
x_n^{D-1} &=& \frac{2\pi M_S}{\abs{a}}\cos(2\pi n u )\cosh(b-a\sigma) \non\\
x_n^D &=& \frac{2\pi M_S}{\abs{a}}\sin(2\pi n u )\cosh(b-a\sigma)
\label{wlinststring}
\ear
(with the remaining coordinates constants).
We take equal signs for $n$ and $a$. Plugging this ansatz into (\ref{eomstring}) one finds
that all the parameters can be determined, namely 

\bear
T_0 &=& \frac{2\pi n}{a} \non\\
b &=& {\rm arctanh} \beta_1 \non\\
a &=& 2\Bigl({\rm arctanh}\beta_1+{\rm arctanh}\beta_2\Bigr) \non\\
N &=& \frac{2\pi M_S}{\abs{a}}\non\\
\label{fixpar}
\ear
This yields the worldsheet action

\bear
S [x^{\mu}] = 2\pi^2M_S^2 \frac{n}{a}
\label{Sstring}
\ear
Noting that $a  = 2\pi\epsilon$
this correctly reproduces the exponents of the Bachas-Porrati formula (\ref{bachasporrati}) in the large $M_S$
limit. The method of calculation is, however, substantially simpler than the ones applied in \cite{burgess}
and \cite{bacpor}. Thus we hope that this worldsheet instanton technique will make it feasible to obtain
string pair creation rates also for certain classes of nonconstant electric fields.
An interesting aspect of the constant field instanton (\ref{wlinststring}) is, that its restriction to each boundary
coincides with the worldline instanton (\ref{instconst}). It is not clear whether this property may extend to more
general fields, but if so it would greatly facilitate finding the corresponding worldsheet instantons.

\end{document}